\documentclass[journal=aamick,manuscript=article]{achemso}
\usepackage[version=4]{mhchem} 
\usepackage{tabularray}
\usepackage{amsmath}
\usepackage{epsfig}
\usepackage{epstopdf}
\usepackage{multirow}
\usepackage{graphicx}
\usepackage{subcaption}
\usepackage{siunitx}

\author{Quach Thi Thai Binh}
\affiliation[VNUHCM]
{Faculty of Physics and Physics Engineering, University of Science, Ho Chi Minh City 700000, Viet Nam}
\alsoaffiliation[VNUHCM]
{Vietnam National University, Ho Chi Minh City 700000, Viet Nam}
\author{Thuan Phuoc}
\affiliation[VNUHCM]
{Faculty of Physics and Physics Engineering, University of Science, Ho Chi Minh City 700000, Viet Nam}
\alsoaffiliation[VNUHCM]
{Vietnam National University, Ho Chi Minh City 700000, Viet Nam}
\author{Xuan Hai}
\affiliation[VNUHCM]
{Faculty of Physics and Physics Engineering, University of Science, Ho Chi Minh City 700000, Viet Nam}
\alsoaffiliation[VNUHCM]
{Vietnam National University, Ho Chi Minh City 700000, Viet Nam}
\author{Thang Bach Phan}
\affiliation[INOMAR]
{Center for Innovative Materials and Architectures (INOMAR)}
\alsoaffiliation[VNUHCM]
{Vietnam National University, Ho Chi Minh City 700000, Viet Nam}
\author{Vu Thi Hanh Thu}
\affiliation[VNUHCM]
{Faculty of Physics and Physics Engineering, University of Science, Ho Chi Minh City 700000, Viet Nam}
\alsoaffiliation[VNUHCM]
{Vietnam National University, Ho Chi Minh City 700000, Viet Nam}
\email{vththu@hcmus.edu.vn}
\author{Nguyen Tuan Hung}
\affiliation[NTU]
{Department of Materials Science and Engineering, National Taiwan University, Taipei 10617, Taiwan}
\alsoaffiliation[VNUHCM]
{Faculty of Physics and Physics Engineering, University of Science, Ho Chi Minh City 700000, Viet Nam}
\alsoaffiliation[VNUHCM]
{Vietnam National University, Ho Chi Minh City 700000, Viet Nam}
\email{nguyenth@ntu.edu.tw}
\title[ML-SERS]
{Rapid Machine Learning–Driven Detection of Pesticides and Dyes Using Raman Spectroscopy}


\usepackage{graphicx}
\usepackage{caption}
\usepackage[table]{xcolor}
\usepackage{tabularx}
\begin{document}

\begin{abstract}
The extensive use of pesticides and synthetic dyes poses critical threats to food safety, human health, and environmental sustainability, necessitating rapid and reliable detection methods. Raman spectroscopy offers molecularly specific fingerprints but suffers from spectral noise, fluorescence background, and band overlap, limiting its real-world applicability. Here, we propose a deep learning framework based on ResNet-18 feature extraction, combined with advanced classifiers, including XGBoost, SVM, and their hybrid integration, to detect pesticides and dyes from Raman spectroscopy, called MLRaman. The MLRaman with the CNN–XGBoost model achieved a predictive accuracy of 97.4\% and a perfect AUC of 1.0, while it with the CNN–SVM model provided competitive results with robust class-wise discrimination. Dimensionality reduction analyses (PCA, t-SNE, UMAP) confirmed the separability of Raman embeddings across 10 analytes, including 7 pesticides and 3 dyes. Finally, we developed a user-friendly Streamlit application for real-time prediction, which successfully identified unseen Raman spectra from our independent experiments and also literature sources, underscoring strong generalization capacity. This study establishes a scalable, practical MLRaman model for multi-residue contaminant monitoring, with significant potential for deployment in food safety and environmental surveillance.
\end{abstract}
\section{Introduction}
The widespread application of pesticides and synthetic dyes has contributed to boosting crop yields ~\cite{Singh2016,Liu2013} and supporting industrial advancement ~\cite{Aigbe2023, Alegbe2024}. Nevertheless, their widespread application also introduces significant risks to food safety ~\cite{Liu2013, Ngegba2022,Guerra2017}; ecological balance ~\cite{Liu2013,Tang2020}; and human well-being ~\cite{Liu2013, Ngegba2022,Guerra2017}.
Pesticides such as carbendazim, carbaryl, thiram, thiabendazole, methyl parathion, cypermethrin, and chlorpyrifos are routinely employed in modern agriculture for crop protection ~\cite{Singh2016,Liu2013,Hu2020}, while synthetic dyes, including rhodamine 6G, rhodamine B, and crystal violet, are commonly used in food processing, textiles, and biomedical industries ~\cite{Aigbe2023,Khan2019,Li2010}. Despite their utility, these substances have been shown to exert profound toxicological impacts: carbendazim and thiabendazole function as endocrine disruptors ~\cite{Rama2014,Ferreira2008,Zheng2024,Mller2014}; carbaryl and methyl parathion trigger neurotoxicity through cholinesterase inhibition ~\cite{Mora-Gutirrez2022,Edwards2005}; thiram contributes to dermatological and neurological disorders ~\cite{Mo2022}; and rhodamine dyes, as well as crystal violet, exhibit genotoxic and carcinogenic properties ~\cite{Priya2024,Kovacic2014,Mani2016}. Residues of these hazardous compounds frequently persist in agricultural produce, soil, and water ~\cite{Singh2016,Rama2014,Priya2024,Mani2016}, thereby creating a pressing need for sensitive, rapid, and reliable detection techniques that can safeguard ecosystems and human health ~\cite{Liu2013,Guerra2017,Peng2021,Zhang2023}.
Among the available analytical techniques, Raman spectroscopy is increasingly recognized as a particularly powerful method for detecting pesticide and dye residues because it is rapid, non-destructive, requires minimal sample preparation, and provides molecularly specific vibrational fingerprints of analytes ~\cite{Petersen2021,He2014,Qu2022}. This makes it highly attractive for real-time and in-situ monitoring of contaminants in complex food and environmental matrices. However, its practical deployment is hindered by a range of technical limitations. Fluorescence background often masks Raman peaks, baseline drift interferes with accurate spectral alignment, and overlapping vibrational bands among structurally similar compounds create significant challenges for reliable discrimination ~\cite{Wei2015,Cai2018}. Furthermore, spectra acquired under field conditions tend to be noisier and less reproducible compared to those collected under controlled laboratory settings, which complicates downstream data analysis and reduces classification accuracy.

Machine learning (ML) methods are being increasingly adapted for materials research due to their ability to analyze complex spectral data~\cite{andrejevic2022machine,hung2024universal}. For example, ML models such as support vector machine classifiers (SVM) and ensemble tree classifiers (RF) have been applied to the detection of thiabendazole and rhodamine 6G dye from the Raman spectra, in some cases exceeding 90\% accuracy on datasets of limited size and analyte diversity~\cite{Liu2025,Yu2025}. Nevertheless, these models still required extensive preprocessing steps, including baseline correction, smoothing, and normalization, to produce reliable results. 
The advent of deep learning has introduced a paradigm shift in Raman spectral analysis. The deep learning models, such as convolutional neural networks (CNNs), can automatically learn hierarchical feature representations directly from raw or minimally processed spectra, thereby reducing dependency on manual preprocessing and improving adaptability across diverse experimental conditions. The CNNs significantly outperformed SVM or RF models in classifying pesticide residues under noisy environments, highlighting their robustness to spectral variability~\cite{Yang2022,Hu2023}. More advanced ML architectures based on CNNs, including ResNet and Inception, have been shown to enhance generalization and discrimination across multi-class spectral datasets, enabling improved performance even in scenarios involving structurally similar compounds~\cite{Xie2023,Hu2022}. In addition, hybrid frameworks that combine CNN-based feature extraction with classical classifiers such as SVM or gradient boosting methods like XGBoost have reported competitive performance~\cite{Wu2023,Taparhudee2024}, further demonstrating the flexibility of CNN-based approaches. 

In this study, we propose an end-to-end ML framework based on the ResNet-18 architecture for the classification of ten hazardous compounds, including seven pesticides (carbendazim, carbaryl, thiram, thiabendazole, methyl parathion, cypermethrin, and chlorpyrifos) and three synthetic dyes (rhodamine 6G, rhodamine B, and crystal violet). The Raman spectral data employed in this work were collected from publicly available datasets reported in prior studies, ensuring both scientific transparency and reproducibility. Instead of relying solely on one-dimensional spectral representations, the Raman spectra were transformed into spectral images, enabling ResNet-18 to leverage its two-dimensional convolutional filters to capture both global spectral patterns and local vibrational features. This design choice exploits the strength of residual connections in mitigating vanishing gradient problems while enabling deeper hierarchical feature learning, thereby enhancing the model’s robustness to noise and variability. The adoption of ResNet-18 is motivated by its well-established balance between computational efficiency and classification performance, making it suitable for both research-oriented applications and potential deployment in field-portable Raman detection systems. In doing so, this study not only extends the scope of Raman-based machine learning applications to encompass a broader set of analytes, including both pesticides and dyes, but also demonstrates the potential of deep residual networks to provide scalable and generalizable solutions for multicompound detection. Ultimately, the proposed approach aims to advance Raman spectroscopy from a powerful laboratory technique to a practical, real-world tool for ensuring food safety, protecting environmental integrity, and safeguarding human health.

\section{Theoretical and Experimental Methods}

\subsection{MLRaman: A machine learning model for detecting pesticides and dyes from Raman spectra}

\begin{figure}
    \centering
    \includegraphics[width=0.95\textwidth]{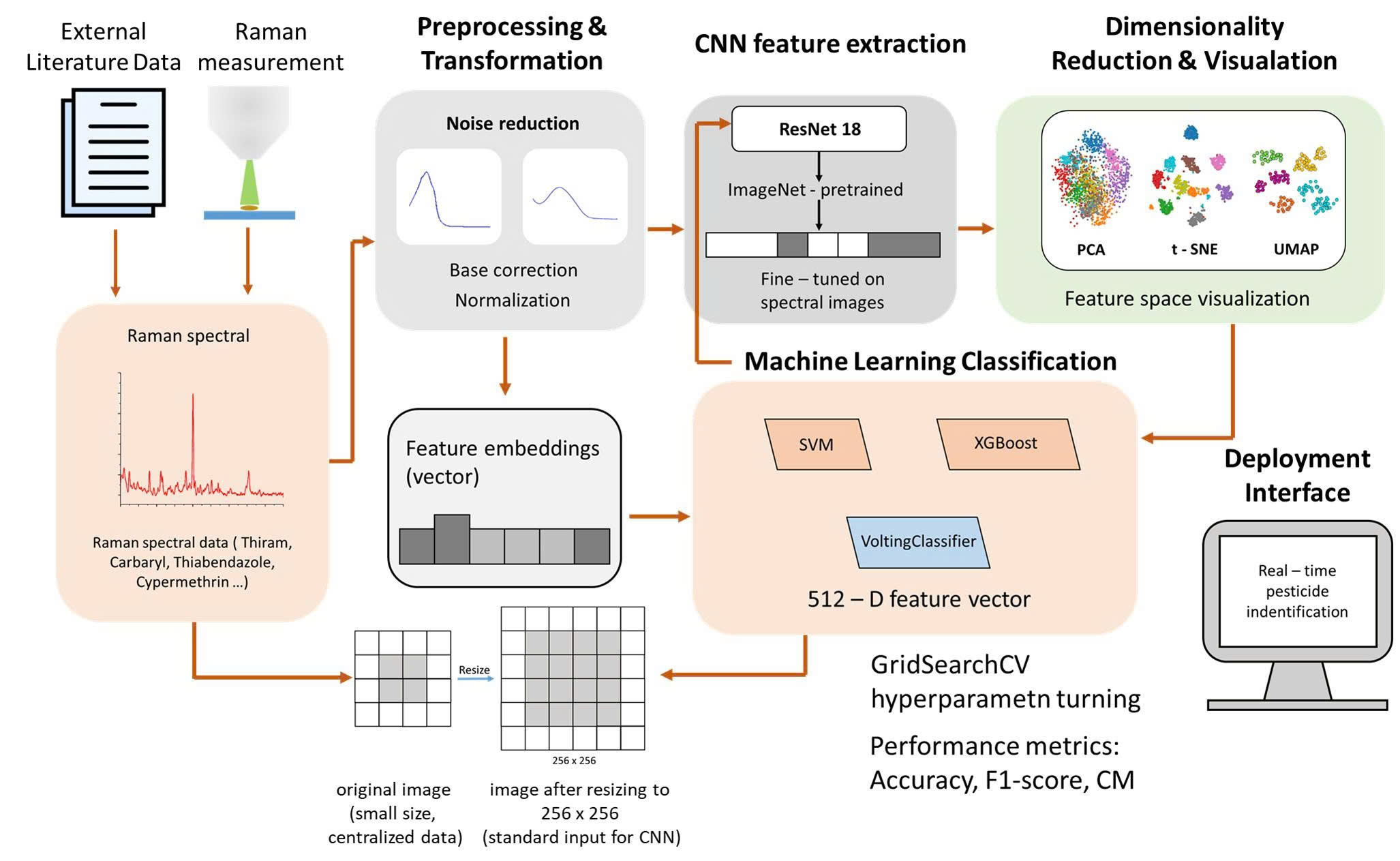} 
    \caption{An integrated pipeline for pesticide identification using Raman spectroscopy and hybrid CNN–machine learning classification. Firstly, the Raman spectral data are collected from literature sources or experimental measurements, which undergo signal preprocessing, including baseline correction and normalization to reduce noise and variability. Then, the processed 1D spectra are converted into 2D image representations and resized uniformly to $256\times 256$ pixels. Next, a ResNet-18 model, pretrained on ImageNet and fine-tuned on spectral images, extracts 512-dimensional deep feature representations. These embeddings are used for both dimensionality reduction (via PCA, t-SNE, and UMAP) to explore class separability, and for classification using support vector machines (SVM), eXtreme Gradient Boosting (XGBoost), and ensemble VotingClassifier models. Classifier optimization is performed using GridSearchCV. Finally, performance is quantitatively evaluated with accuracy, F1-score, and a confusion matrix.}
    \label{fig:MLRaman}
\end{figure}

In Fig.~\ref{fig:MLRaman}, we illustrate the MLRaman framework used in this study. The framework begins with the Raman spectral dataset and data preprocessing, then moves on to feature extraction with ResNet18, followed by dimensionality reduction and classification using several ML models, including SVM, XGBoost, and VotingClassifier. The performance of the MLRaman framework is measured with accuracy, F1-score, and a confusion matrix. A user interface has also been developed to allow for practical use of the model for the rapid detection of pesticides and dyes from Raman spectra. All code and the user interface application are available for download on the GitHub page: \url{https://github.com/nguyen-group/MLRaman}.

\subsection{Dataset and preprocessing}
This study employs a dataset of Raman spectral data corresponding to ten pesticide-related compounds: carbendazim (CBZ), carbaryl (CR), thiram (TMTD), thiabendazole (TBZ), rhodamine 6G (R6G), rhodamine B (RB), crystal violet (CV), methyl parathion (MP), cypermethrin (CYP), and chlorpyrifos (CPF). The raw Raman spectra are obtained from publicly available databases, published literature, and in-house measurements. Each spectrum is stored as intensity versus wavenumber data in the conventional one-dimensional (1D) format. The dataset can be found on the GitHub page: \url{https://github.com/nguyen-group/MLRaman/data}.

To enable image-based deep learning, raw spectra are baseline-corrected, normalized, and interpolated, then converted into two-dimensional (2D) pseudo-color images. This conversion preserves molecular vibrational fingerprints while making them compatible with convolutional neural networks (CNNs). All images are resized to $256 \times 256$ pixels (see Fig.~\ref{fig:MLRaman}), ensuring compatibility with standard CNN backbones such as ResNet18 and EfficientNet. Intensities are rescaled to the range $[0, 1]$ to stabilize training, and mild augmentations (random rotation, color jitter, and erasing) are applied to improve model robustness.

\begin{figure}[ht!]
    \centering
    \includegraphics[width=0.95\textwidth]{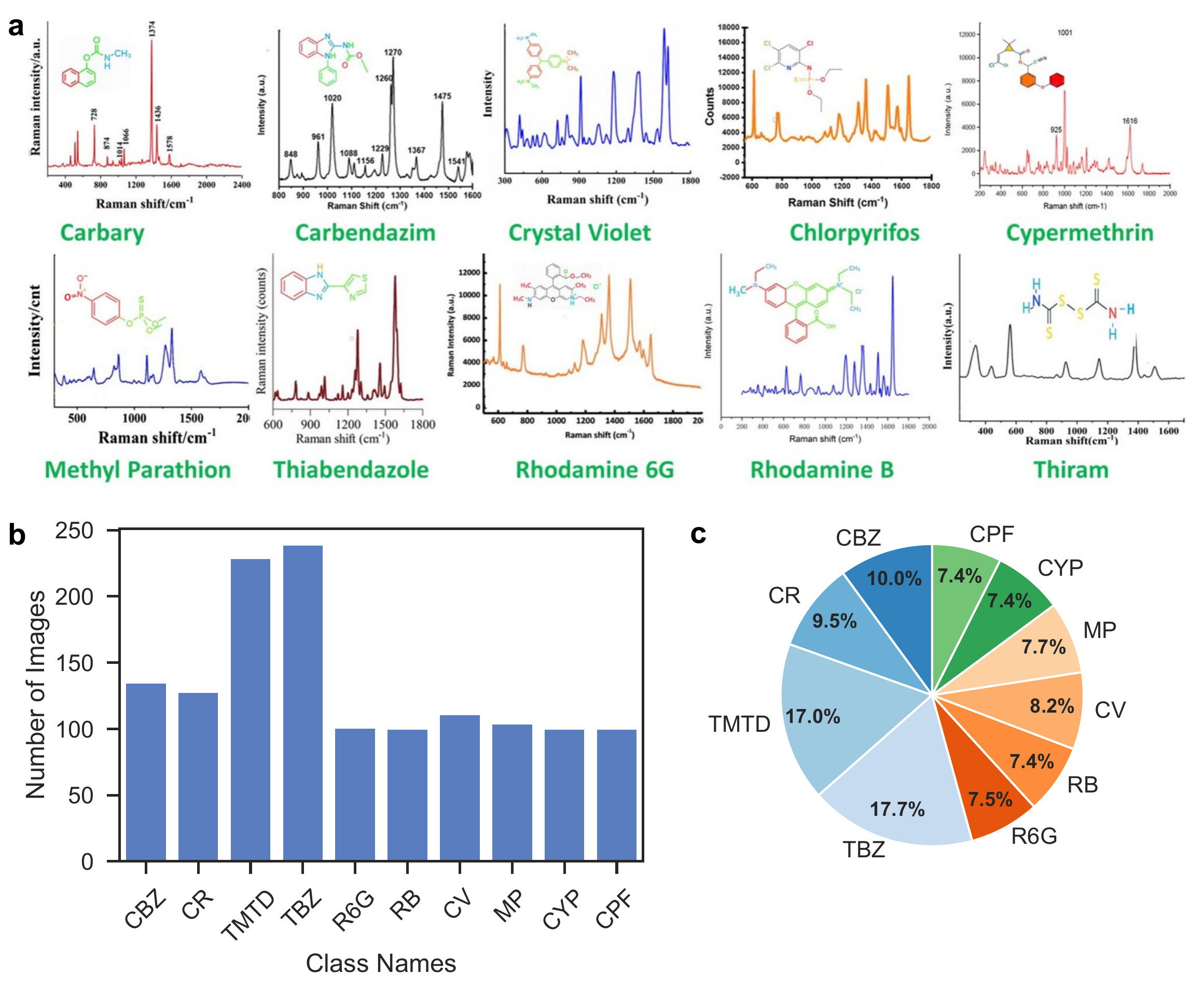} 
    \caption{Raman spectra dataset for 10 pesticides and dyes. (a) The Raman spectra of the 10 compounds, including carbendazim (CBZ), carbaryl (CR), thiram (TMTD), thiabendazole (TBZ), rhodamine 6G (R6G), rhodamine B (RB), crystal violet (CV), methyl parathion (MP), cypermethrin (CYP), and chlorpyrifos (CPF), and their molecular formulas. (b) Bar chart showing the distribution of Raman spectral images across 10 chemical classes. The dataset shows class imbalance, with TBZ and TMTD having the largest sample counts. (c) Corresponding pie chart illustrating the proportion of each class within the dataset, reflecting a relatively skewed but comprehensive representation of the target analytes.}
    \label{fig:dataset}
\end{figure}

The dataset exhibited a relatively balanced distribution across classes, as shown in Fig.~\ref{fig:dataset}. {TMTD} ($\approx$ 17.7 \%) and {TBZ} ($\approx$ 17.0 \%) are the largest categories, while smaller ones such as CPF and MP still contained sufficient samples to avoid severe imbalance. This distribution supports fair training and reliable evaluation across all target pesticides.

\subsection{CNN-based feature extraction using ResNet}
In the MLRaman, we use ResNet18 to extract discriminative spectral–spatial features from Raman pseudo-color images. Instead of training from scratch, the ResNet18 is initialized with ImageNet-pretrained weights and fine-tuned on the spectral dataset~\cite{deng2009imagenet}. To obtain generalizable representations, features are collected from the penultimate layer, just before the classification head. These embeddings capture abstract but discriminative information, serving as compact input for subsequent ML classifiers.

\subsection{Machine learning classifiers}
Extracted CNN features are used to train several ML classifiers, including SVM, XGBoost, and VotingClassifier (combining SVM and XGBoost). 

The hyperparameters are optimized using GridSearchCV, which systematically evaluates candidate parameter values. 
For the SVM, the penalty parameter C (1, 2, 4, 8), the kernel type (a linear and radial basis function (RBF) kernels), and kernel coefficient $\gamma$ (0.1, 0.01, 0.001) are tuned. For the XGBoost, the parameters are the number of estimators (100-500), the learning rate (0.01, 0.05, 0.1), and the maximum depth of the tree (3-10). For the ensemble VotingClassifier, hard and soft voting strategies are evaluated.

All models are validated using 5-fold cross-validation, a widely adopted method that mitigates overfitting by averaging performance across multiple data partitions. Performance is evaluated by accuracy, F1-score, and confusion matrices.

\subsection{Dimensionality reduction and visualization}
To interpret the learned representations, three dimensionality reduction methods, including principal component analysis (PCA), t-distributed stochastic neighbor embedding (t-SNE), and uniform manifold approximation and projection (UMAP) are applied in the MLRaman. Each method projects high-dimensional CNN embeddings into a 2D space, facilitating the visualization of pesticide category separability. Side-by-side comparisons of the projections highlight clustering clarity and provide insights into the latent feature structure.

\subsection{Experimental Raman measurement}
Raman measurements are performed using a Horiba XploRA PLUS spectrometer with a 532 nm excitation laser, focused through a 50$\times$ long-working-distance objective lens to a spot size of approximately 2 $\mu$m. The Raman signals are collected at a laser power of 0.1 mW with an acquisition time of 5 s, ensuring both effective excitation and reliable signal detection. In this work, crystal violet (CV, 97\% purity, Merck) and rhodamine 6G (R6G, 99\% purity, Sigma-Aldrich) are used as analyte solutions. The target molecules are dissolved in deionized water at concentrations ranging from 10$^{-3}$ M to 10$^{-4}$ M. For Raman measurements, 20 $\mu$L of each solution is deposited onto a glass substrate, dried at room temperature, and then subjected to Raman analysis.

\section{Results and Discussion}  

\subsection{Raman preprocessing}

\begin{figure}[t]
       \centering
        \includegraphics[width=0.6\textwidth]{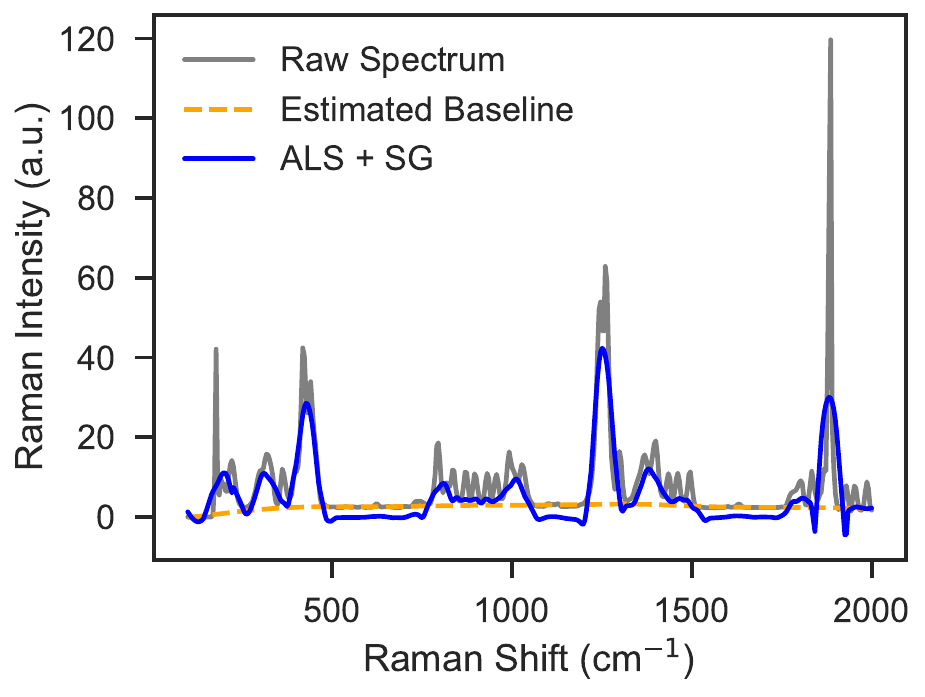} 
         \caption{Comparison between the raw Raman spectrum of thiram (gray), the estimated baseline via ALS (orange dashed), and the preprocessed spectrum after baseline removal and SG smoothing (blue). The pipeline suppresses baseline drift, enhances peak clarity, and reduces high-frequency noise, yielding standardized spectral inputs for machine-learning models.}
        \label{fig:Raman-Thiram}
\end{figure}

To evaluate the impact of the preprocessing pipeline on spectral quality, we compare and process Raman spectra of Thiram, as shown in Fig.~\ref{fig:Raman-Thiram}. Thiram is selected as a representative compound because its raw spectrum clearly shows baseline drift and fluorescence background, making it suitable for demonstrating the effect of preprocessing.
The raw spectrum (gray) shows a slowly varying baseline and background fluctuations that obscure several characteristic vibrational peaks. The estimated baseline (orange dashed line), which is obtained by asymmetric least squares (ALS) fitting, captures this broad low-frequency trend. After baseline subtraction and subsequent Savitzky–Golay (SG) smoothing, the preprocessed spectrum (blue) shows a more flattened baseline, and peaks are more pronounced than in the raw data.

Three main improvements can be observed: (i) baseline suppression, which eliminates background offsets and prevents the model from learning irrelevant intensity variations; (ii) peak enhancement, where diagnostic Raman bands emerge more distinctly, facilitating feature extraction; and (iii) noise reduction, where high-frequency fluctuations are smoothed while preserving both peak position and relative intensity.
Although Figure~\ref{fig:Raman-Thiram} illustrates the case of thiram, similar improvements are consistently observed across the spectra of other compounds in the dataset. This confirms that the ALS+SG preprocessing pipeline provides cleaner, more standardized inputs for downstream ML tasks, preserving discriminative vibrational information while minimizing irrelevant artifacts.

\subsection{Baseline CNN training performance}
\begin{figure}[t]
    \centering
    \includegraphics[width=0.95\textwidth]{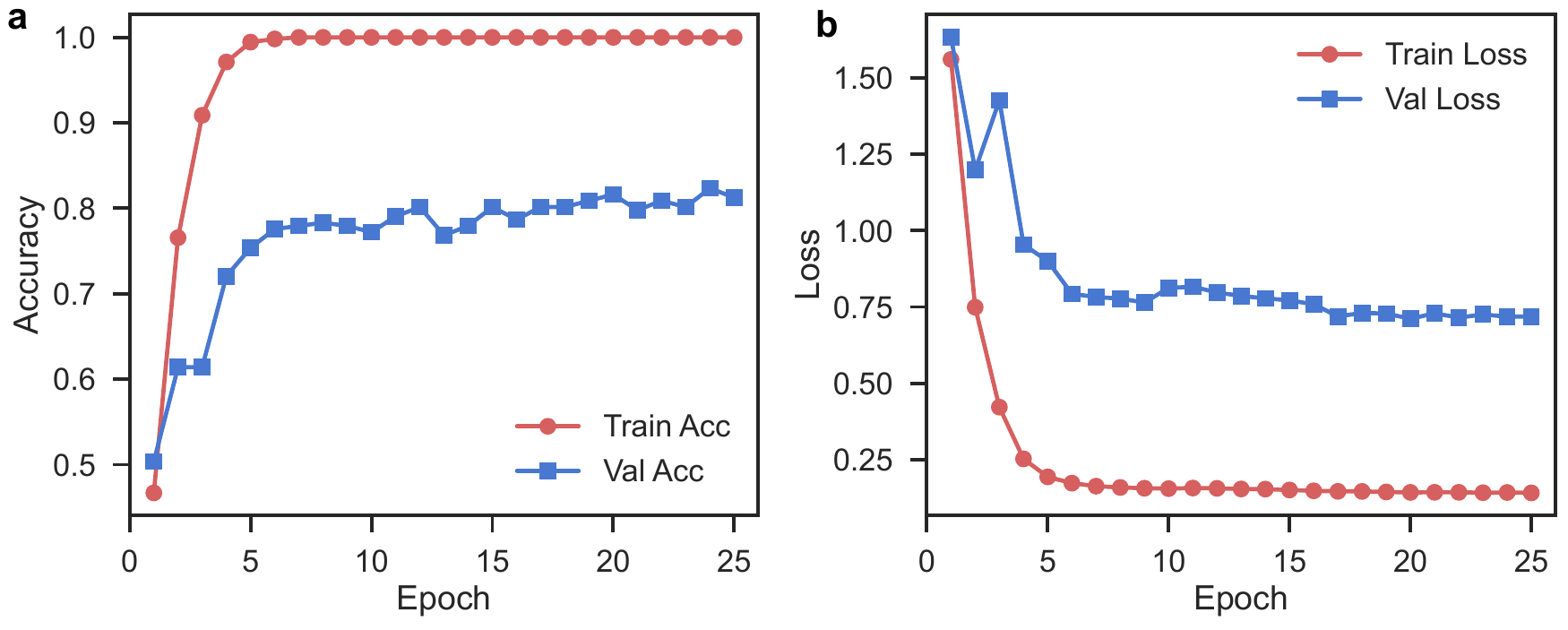} 
    \caption{Training and validation performance of the ResNet18 model over 30 epochs, showing decreasing accuracy (a) and loss (b) history with validation accuracy plateauing at 80\%.}
    \label{fig:loss}
\end{figure}

The proposed ResNet18-based architecture showed apparent convergence and reliable generalization. As shown in Fig.~\ref{fig:loss}, training loss decreased substantially from 2.19 to 0.35, while validation loss decreased steadily from 1.70 to 0.88 over 25 epochs. Concurrently, training accuracy improved from 46.7\% to 100.0\%, while validation accuracy increased steadily from 50.4\% to a peak of 82.3\% at epoch 24 and converged around 80\%. The rapid initial gains followed by a stable plateau indicate that the network successfully learned discriminative representations that generalized robustly to unseen samples. Importantly, the stability of validation performance despite near-perfect training accuracy reflects a well-regularized learning process that avoids severe overfitting. 
Taken together, these findings confirm the effectiveness of CNN-based transfer learning for this classification task and establish a strong foundation for further refinement to surpass the current 80\% benchmark.

\subsection{Performance of the CNN-XGBoost classification}

\begin{figure}[t]
    \centering
    \includegraphics[width=0.90\textwidth]{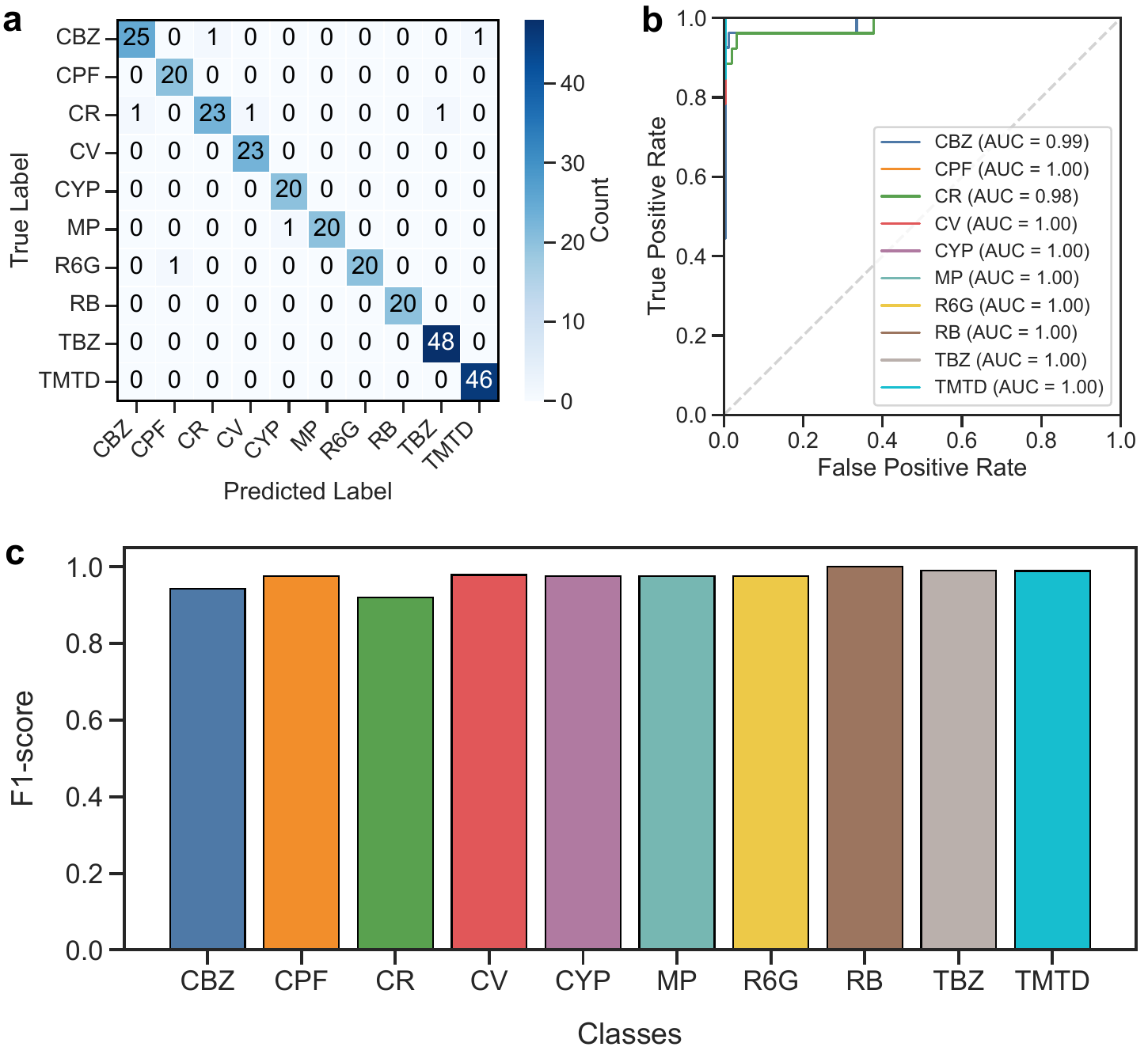} 
    \caption{Performance evaluation of XGBoost model with the CNN features and the PCA dimensionality reduction, including (a) confusion matrix, (b) ROC curves, and (c) per-class F1-scores, for the validation dataset.}
    \label{fig:xgb}
\end{figure}

Compared with conventional strategies extensively reported in Raman spectral analysis, such as SVM or linear discriminant analysis (LDA)-based classifiers, our framework demonstrates a clear advantage. Traditional methods typically falter in the presence of overlapping vibrational bands and experimental noise, often plateauing at 80–85\% precision~\cite{liu2017deep, huang2022raman}. Even CNN-only approaches, as reflected in both previous studies~\cite{huang2022raman} and our baseline experiments, fail to surpass 85\% validation accuracy under identical conditions. In stark contrast, the CNN–XGBoost hybrid not only outperforms these baselines but also establishes a robust, stable, and highly generalizable decision boundary.

In our framework, the hybrid CNN–XGBoost model shows a substantial performance leap, achieving an overall classification accuracy of 97.4\%. As shown in Fig.~\ref{fig:xgb}a, the confusion matrix summarizes the performance of a classification model for the validation dataset. The number of validation datasets for CBZ, CPF, CR, CV, CYP, MP, R6G, RB, TBZ, and TMTD is 27, 20, 26, 23, 20, 21, 21, 20, 48, and 46, respectively. Compared with the counts in the confusion matrix, we can see that all samples in the CPF, CV, CYP, RB, and TBZ classes are correctly predicted, while the MP and R6G classes each have one failed sample, the CBZ class has two, and the CR class has three. Moreover, the area under the curve (AUC) for the validation dataset shows perfect (AUC $=1.00$) across all classes except CBZ (AUC $=0.99$) and CR (AUC $=0.98$), as shown in Fig.~\ref{fig:xgb}b, indicating the near-ideal separability of Raman fingerprints. The F1-score, a harmonic mean of precision and recall, is also used to evaluate the performance of classification models, where an F1 score reaches its best value of 1 and its worst value of 0. As shown in Fig.~\ref{fig:xgb}c, the CBZ and CR are the only cases with a slightly lower F1-score (0.94 and 0.92), attributable to minor recall fluctuations. However, such deviations are isolated and negligible in terms of impact on overall predictive performance.

The high-performance of CNN+XGBoost comes from the complementary strengths of the two models: CNN provides a hierarchical feature representation that encodes subtle spectral variations~\cite{liu2017deep}, while XGBoost delivers a powerful ensemble-based decision mechanism capable of capturing non-linear feature interactions and mitigating class imbalance~\cite{chen2016xgboost,cheng2024machine}. This synergy effectively overcomes the limitations of both conventional classifiers and deep learning models when used in isolation. Therefore, our model sets a new benchmark for Raman-based chemical classification. The combination of high overall accuracy, excellent class consistency, and almost perfect AUC underscores the strong potential of the proposed framework for real-world deployment, which includes rapid pesticide detection, dye contamination monitoring, and broader applications in environmental and food safety surveillance.

\subsection{Performance of the CNN-SVM classification}

\begin{figure}[t]
\centering
\includegraphics[width=0.90\textwidth]{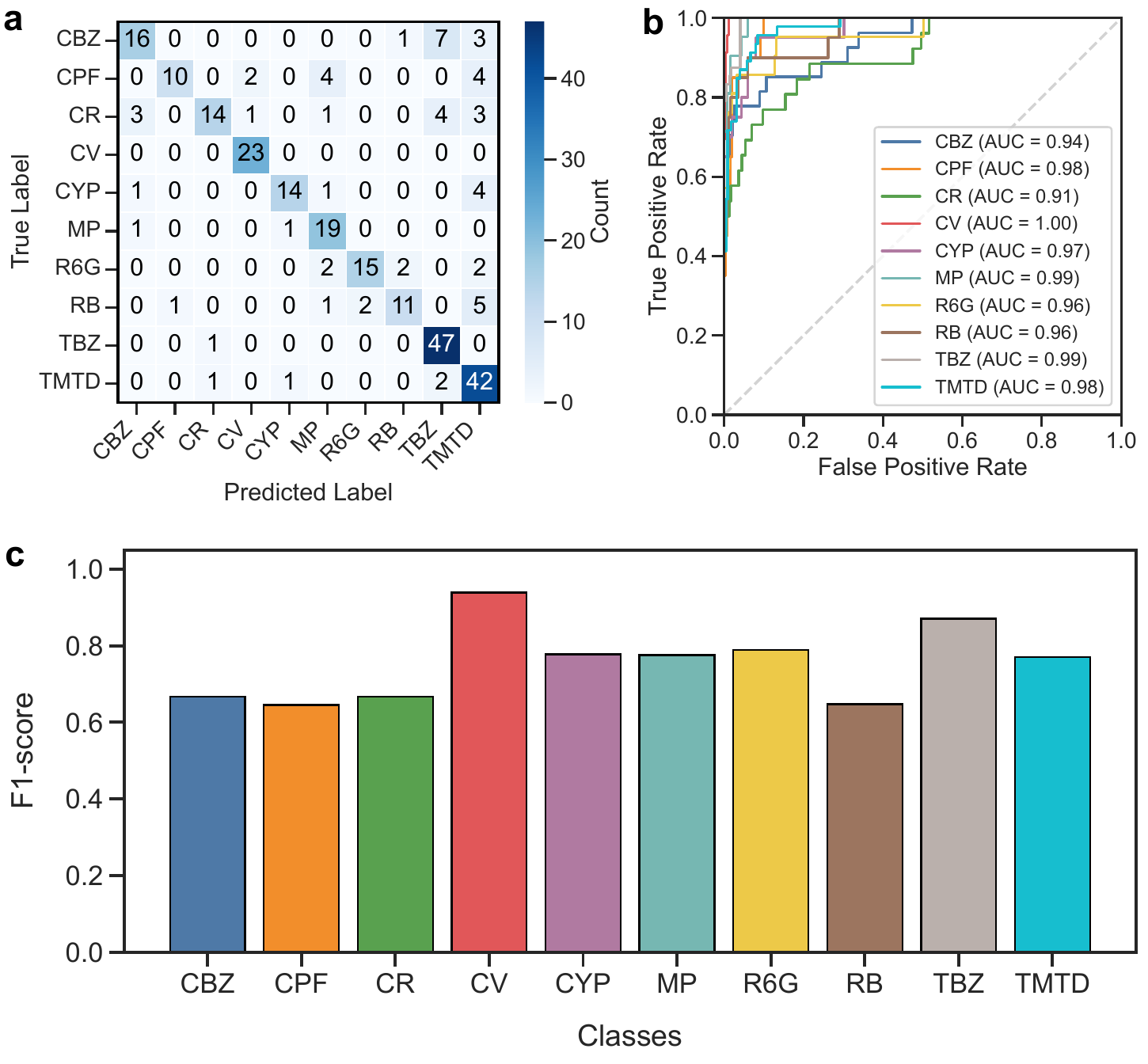} 
\caption{Performance evaluation of the SVM model with the CNN features and the PCA dimensionality reduction, including (a) confusion matrix, (b) ROC curves, and (c) per-class F1-scores, for the validation dataset.}
\label{fig:svm}
\end{figure}

To benchmark against the XGBoost model, we also perform SVM and VotingClassifier (combining SVM and XGBoost) on the CNN features and PCA-dimensionality reduction. In Fig.~\ref{fig:svm}, we show the performance of the GNN+SVM model for the validation dataset. The confusion matrix (Fig.~\ref{fig:svm}a) shows that most classes are correctly assigned to their respective categories, with only a few misclassifications observed in cases of spectrally similar compounds. This indicates that the SVM effectively captures fine vibrational distinctions that are often obscured by baseline shifts or overlapping peaks. The performance analysis of the SVM model is also presented as ROC curves and F1-scores. Overall, the GNN+SVM model is less efficient than the GNN+XGBoost model in Raman classification. However, it is noted that the SVM performance is sensitive to kernel selection and hyperparameter optimization, which may limit its scalability to larger, more heterogeneous datasets. By using GridSearchCV~\cite{pedregosa2011scikit} to automatically optimize hyperparameters, we can improve the classification accuracy of GNN+SVM from 77.6\% to 96.3\%. Although it is still below GNN+XGBoost (97.4\%), this complementarity suggests that integrating both approaches within a hybrid framework could enhance generalization performance, thereby advancing Raman-based contaminant detection toward real-world applications.

\begin{figure}[ht!]
\centering
\includegraphics[width=0.90\textwidth]{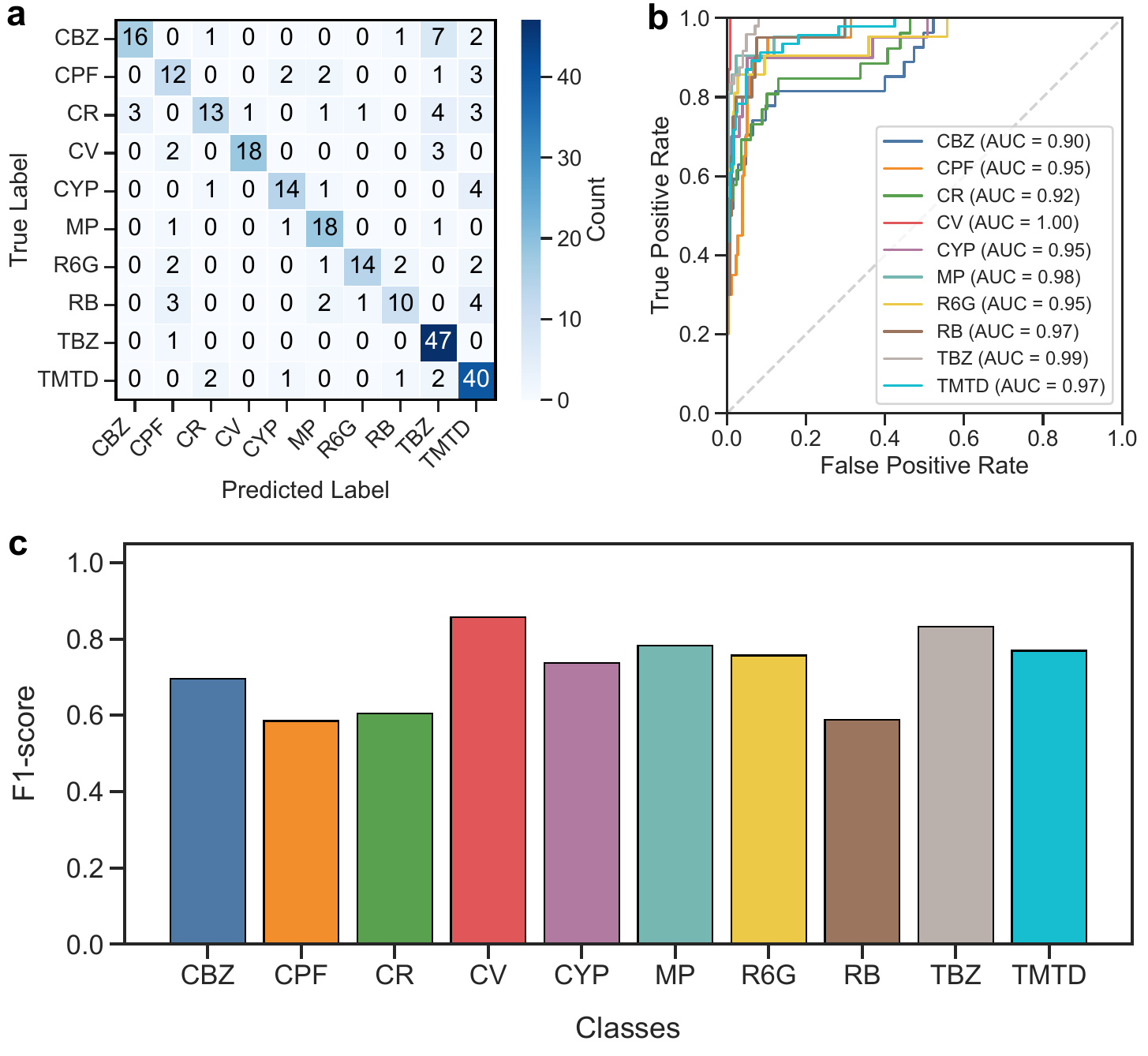} 
\caption{Ensemble classification performance using the VotingClassifier framework (combining SVM and XGBoost) with the CNN features and the PCA dimensionality reduction, including (a) confusion matrix, (b) ROC curves, and (c) per-class F1-scores, for the validation dataset.}
\label{fig:voting}
\end{figure}

\subsection{Ensemble learning with VotingClassifier}
Next, we will combine both SVM and XGBoost within a VotingClassifier framework. This ensemble strategy balances the margin-maximization property of SVM with the gradient-boosting capability of XGBoost, thereby reducing misclassification errors across closely related spectra. In Fig.~\ref{fig:voting}, we show the performance of the VotingClassifier framework using the GNN features and the PCA dimensionality reduction for the validation dataset. The confusion matrix, ROC curves, and per-class F1-scores are quite similar to those of the SVM model (see Fig.~\ref{fig:svm}). Although ensemble learning with VotingClassifier is expected to improve performance in Raman spectral classification, the VotingClassifier achieves a classification accuracy of 76.1\%, which is lower than that of XGBoost (97.4\%) and SVM (77.6\%). This might be due to imbalanced or noisy Raman data. In this case, the SVM and XGB disagree frequently, leading to an unstable ensemble.

\subsection{Dimensionality reduction and feature visualization}

\begin{figure}[t]
    \centering
    \includegraphics[width=0.95\textwidth]{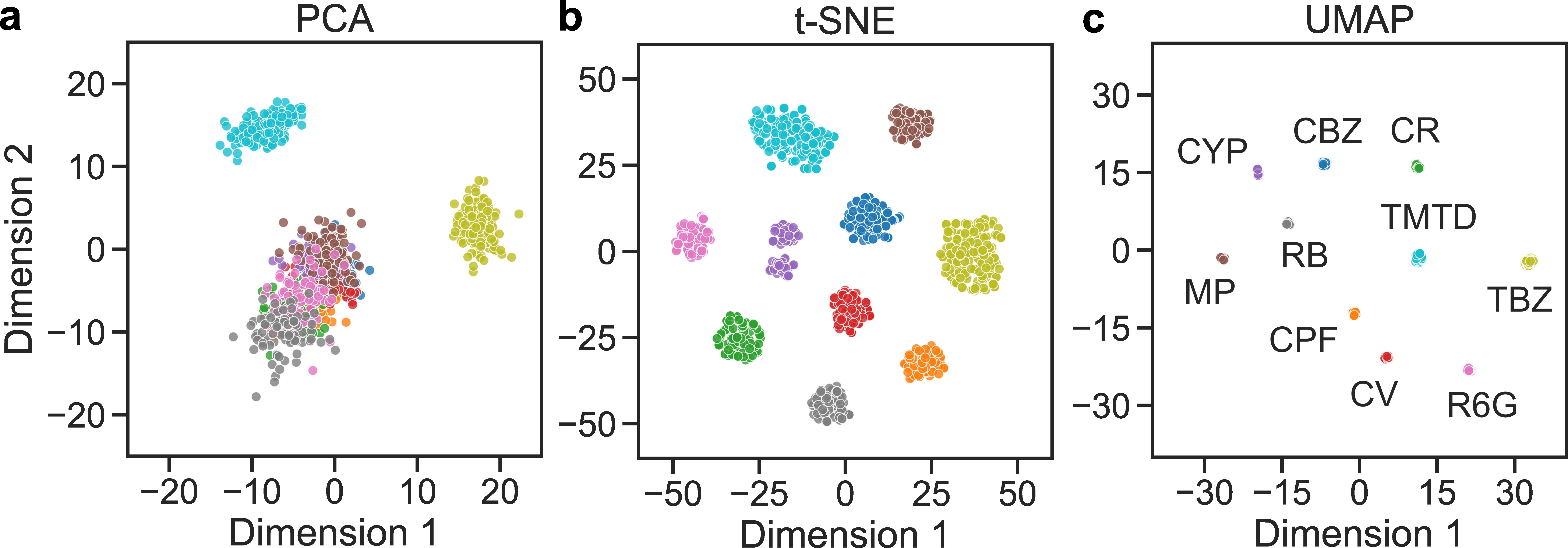} 
    \caption{Dimensionality reduction and visualization of CNN feature embeddings using (a) PCA, (b) t-SNE, and (c) UMAP. Color dots indicate 10 pesticides and dyes, including carbendazim (CBZ), carbaryl (CR), thiram (TMTD), thiabendazole (TBZ), rhodamine 6G (R6G), rhodamine B (RB), crystal violet (CV), methyl parathion (MP), cypermethrin (CYP), and chlorpyrifos (CPF).}
    \label{fig:pca}
\end{figure}

In Fig.~\ref{fig:pca}, we use dimensionality reduction methods to visualize the CNN feature space in two dimensions to further elucidate why such improvements occur. These visualizations provide valuable insight into how pesticides and dyes are distributed in the embedding space and where classification challenges arise.

As shown in Fig.~\ref{fig:pca}a, for the PCA method, linear projection preserves the global variance of features but fails to fully separate chemically similar pesticides such as carbaryl (CR) and carbendazim (CBZ), while it well separates the thiabendazole (TBZ) and thiram (TMTD). This may be due to the dataset distribution (see Fig.~\ref{fig:dataset}), where the number of samples in the TBZ and TMTD classes is significantly higher than that in the other classes.   This also explains why the TBZ and TMTD classes are well classified in all models (XGBoost, SVM, and VotingClassifier).

In contrast, the t-SNE method highlights the ability of CNN embeddings to form distinct clusters for most classes, as shown in Fig.~\ref{fig:pca}b, reflecting their discriminative power. The UMAP method shows the clearest separation of classes compared with PCA and t-SNT methods, as shown in Fig.~\ref{fig:pca}c. We use all dimensionality reduction methods for the CNN feature in the CNN+SVM model to test the performance. We find a slight improvement in classification accuracy: 77.6\% for PCA and 79.8\% for both t-SNE and UMAP. This improvement is due to the better feature separation ability of t-SNE and UMAP compared to PCA. Nevertheless, training time also increased significantly because t-SNE and UMAP are non-linear methods, whereas PCA is linear. Therefore, the PCA is a good choice for computational efficiency.

\begin{figure}[ht!]
    \centering
    \includegraphics[width=0.90\textwidth]{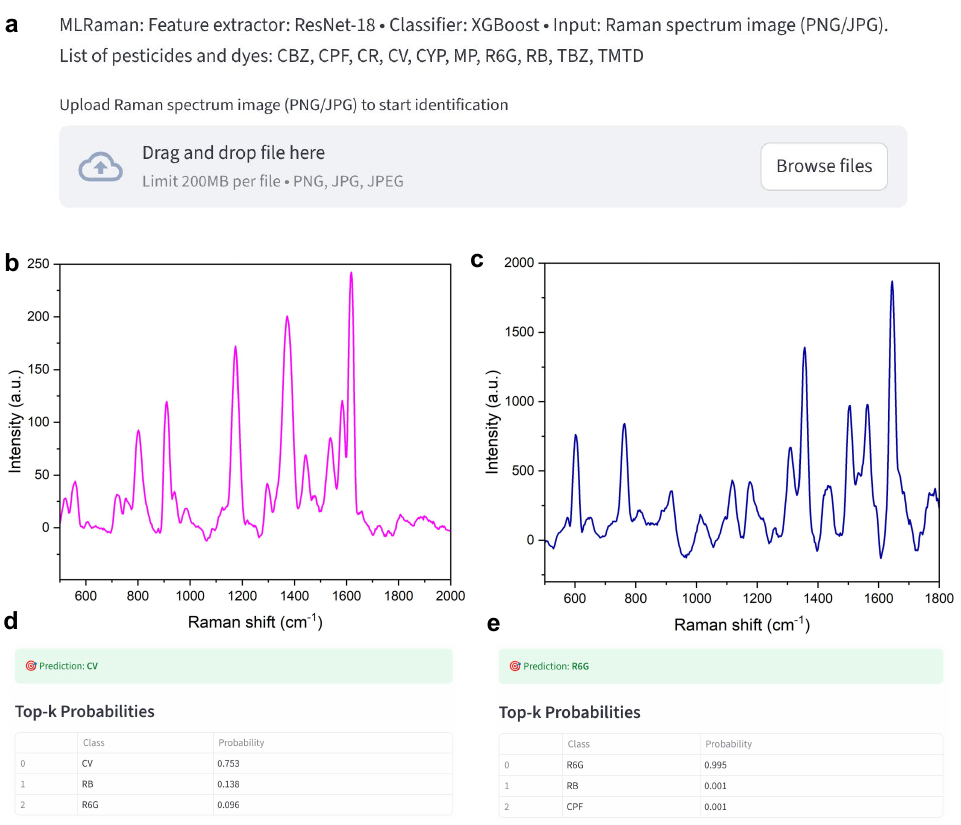} 
    \caption{A user-friendly Streamlit application based on the MLRaman model with GNN+XGBoost. (a) The app only requires uploading a Raman spectrum from a list of 10 pesticides and dyes, including carbendazim (CBZ), carbaryl (CR), thiram (TMTD), thiabendazole (TBZ), rhodamine 6G (R6G), rhodamine B (RB), crystal violet (CV), methyl parathion (MP), cypermethrin (CYP), and chlorpyrifos (CPF). (b) Unseen Raman spectrum of CV and (c) R6G. (d) and (e) show the top three prediction probabilities for CV and R6G, respectively, which are the correct predictions for the Raman spectra in (b) and (c), respectively.}
    \label{fig:app}
\end{figure}

\section{App-based External Test on Unseen Raman Spectra}

To assess deployability beyond curated datasets, we conducted a prospective, app-based evaluation using the Streamlit interface for the MLRaman with the GNN+XGBoost model. The application accepts a spectral image, automatically crops the plot area (binary thresholding and bounding-box extraction to remove titles/axes), rescales it to the training resolution, and applies the same normalization used during model development before forwarding the image to the MLRaman classifier. We deliberately used unseen spectra—distinct from all training/validation files—sourced from our experimental Raman spectra for two samples: the crystal violet (CV) and the rhodamine 6G (R6G).

This app only requires uploading a Raman spectrum, as shown in Fig~\ref{fig:app}a, then the app returns correct top-1 predictions for every case. As shown in Fig~\ref{fig:app}b-e, the MLRaman can predict and correct the CV and R6G with the probability of 0.753 and 0.995, respectively. This predictive ability is remarkable because the CV and R6G are dyes with partially overlapping Raman bands, making them one of the most confusable pairs in routine analyses. Correct assignment of unseen images indicates that the learned representation captures fine-grained, class-specific motifs rather than overfitting to dataset-specific artifacts (e.g., axis grids, font styles). Therefore, the cropping-plus-normalization stage combined with residual feature extraction confers robustness to baseline drift and moderate illumination/format variability.

The MLRaman app can be extended to analyze any Raman spectra, not just those related to 10 specific pesticides and dyes. However, a significant challenge we face is obtaining reliable data sources and addressing data imbalance issues. Our goal in developing the MLRaman is to enhance our datasets through laboratory experiments and theoretical simulations. To achieve this, autonomous experiments powered by active learning and artificial intelligence, along with high-throughput computational methods, could be solutions in the near future.

\section{CONCLUSION}
This work presented a machine–learning–based framework for Raman spectral classification of 10 hazardous pesticides and dyes. By leveraging ResNet-18 for feature extraction and coupling its embeddings with advanced classifiers, particularly XGBoost and SVM, we achieved consistently high accuracy (97,4\%) and near-perfect AUC scores. Dimensionality reduction analyses further confirmed the discriminative power of CNN-derived embeddings. A user-friendly Streamlit application successfully identified external spectra, including independent experimental Raman spectra of crystal violet and rhodamine 6G, demonstrating strong generalization to out-of-source data. These results highlight the promise of integrating CNN feature learning with ensemble models for real-time Raman-based contaminant detection, paving the way toward practical deployment in food safety and environmental monitoring.

\begin{acknowledgement}
This work was financially supported by Vietnam National University Ho Chi Minh City (NCM2024-50-01).
\end{acknowledgement}




\providecommand{\latin}[1]{#1}
\makeatletter
\providecommand{\doi}
  {\begingroup\let\do\@makeother\dospecials
  \catcode`\{=1 \catcode`\}=2 \doi@aux}
\providecommand{\doi@aux}[1]{\endgroup\texttt{#1}}
\makeatother
\providecommand*\mcitethebibliography{\thebibliography}
\csname @ifundefined\endcsname{endmcitethebibliography}  {\let\endmcitethebibliography\endthebibliography}{}

\end{document}